\documentclass[aps,pra,twocolumn]{revtex4-1}
\usepackage{amsmath}
\usepackage{amssymb}
\usepackage{graphicx}
\usepackage{multirow}
\usepackage{amsfonts}
\providecommand{\U}[1]{\protect\rule{.1in}{.1in}}
\newcommand{\abs}[1]{\lvert #1 \rvert}
\begin{document}
\title{Magnetic spiral induced by strong correlations in MnAu$_{2}$}
\author{J.~K.~Glasbrenner}
\affiliation{National Research Council/Naval Research Laboratory, Washington,
DC 20375, USA}
\author{K.~M.~Bussmann}
\affiliation{Naval Research Laboratory, Washington, DC 20375, USA}
\author{I.~I.~Mazin}
\affiliation{Naval Research Laboratory, Washington, DC 20375, USA}
\date{\today}

\begin{abstract}
The compound MnAu$_2$ is one of the oldest known spin-spiral materials, yet the
nature of the spiral state is still not clear. The spiral cannot be explained
via relativistic effects due to the short pitch of the spiral and the weakness
of the spin-orbit interaction in Mn, and another common mechanism, nesting, is
ruled out as direct calculations show no features at the relevant wave vector.
We propose that the spiral state is induced by a competition between the
short-range antiferromagnetic exchange and a long-range interaction induced by
the polarization of Au bands, similar to double exchange. We find that, contrary
to earlier reports, the ground state in standard density functional theory is
ferromagnetic, \emph{i.e.}, the latter interaction dominates. However, an
accounting for Coulomb correlations via a Hubbard $U$ suppresses the
Schrieffer-Wolff type $s-d$ magnetic interaction between Mn and Au faster than
the superexchange interaction, favoring a spin-spiral state. For realistic
values of $U$ the resulting spiral wave vector is in close agreement with
experiment.
\end{abstract}

\maketitle

\section{Introduction}
\label{sect-intro}

The magnetic spiral is a type of noncollinear magnetic ordering in materials in
which the localized moments form a screw-type pattern about an axis. Since the
discovery of the first spin spiral in MnAu$_{2}$ in 1956 \cite{Meyer1956:JPR},
the origins of such spirals have been subject of intensive research. A number of
mechanisms have been discussed. In the local moment Heisenberg-exchange picture,
a natural source of spirals is magnetic frustration of the nearest and
next-nearest neighbor exchange parameters, for example, when $J_{1} \neq 0$
(ferro- or antiferromagnetic) and $J_{2} < 0$ (antiferromagnetic). If the
next-nearest neighbor exchange is large enough, such that $\abs{J_{1}} < 4
\abs{J_{2}}$, then the mean-field solution yields a spiral ground state
\cite{Enz1961:JAP,Nagamiya1962:PTP}. This mechanism is applicable in both
localized and itinerant electron systems, although in the latter one may expect
long range interactions to play a role. For an itinerant system, the structure
of the electronic response in reciprocal space captures the role of such
interactions. If there is a maximum in the spin susceptibility at a particular
wave vector $\mathbf{q}$, then this generates a spin density wave with the same
wave vector, which can take the form of a spin spiral. In principle, a kink (as
in the uniform 2D electron gas) or even a derivative singularity (as in the 3D
electron gas) can induce an oscillatory interaction
(Ruderman-Kittel-Kasuya-Yosida, or RKKY interaction) in real space, which can
also encourage a spiral formation. Contrary to a common misconception, this
effect does not require electron bands crossing the Fermi surface (the
interaction is defined by the real, not imaginary, part of susceptibility), but
it does become weaker as the excitation gap grows. Finally, geometric
frustration can also encourage noncollinearity and helical ordering in
materials.

The Dzyaloshinskii-Moriya (DM) interaction
\cite{Moriya1960:PR,Dzayloshinskii1958:JPCS}, a relativistic effect that occurs
in materials without an inversion center, has attracted a substantial amount of
interest and can lead to moment canting or spiraling. This interaction becomes
more important in materials with heavy elements, such as the rare earth series.
Relativistic effects are important in such materials, although more than one
mechanism is often in play. For example, the spiral phases of the heavy
rare-earth metals Tb-Tm \cite{Jensen1991} are also understood to be due to
nesting \cite{Evenson1968:PRL,Dugdale1997:PRL,Nordstrom2000:EPL}.

Spiral ordering is not restricted to materials with heavy magnetic ions. Spirals
also form in lighter transition metal materials with an intriguing range of
spiral vectors $\mathbf{q}$. The DM mechanism may be operative in some of these
materials, but in such systems the weak relativistic effects induce spirals with
long wavelengths as in MnSi
\cite{Bak1980:JPC,Nakanishi1980:SSC,Kataoka1981:JPSJ}. Shorter wavelength
spirals are also common, such as in the magnetically frustrated spinel chromites
ZnCr$_{2}$Se$_{4}$ \cite{Plumier1966:JPF} and CdCr$_{2}$O$_{4}$
\cite{Chung2005:PRL}. An interesting example is FeAs, featuring an
incommensurate spiral with a period of 20 Fe layers
\cite{Selte1973:ACS,Segawa2009:JPSJ}. This material is a good metal, so one may
think that the conduction electrons mediate an oscillatory interaction via
nesting or the classical 3D RKKY mechanism. However, the search for a nesting
vector or features in the noninteracting susceptibility that match the spin
spiral vector were unsuccessful \cite{Parker2011:PRB}. But, as in the cited
example of the 3D electron gas, oscillatory interactions may manifest themselves
even without such peaks, and so one cannot rule out this mechanism in total
without a full calculation of spin susceptibility in real space.

The material MnAu$_{2}$, as stated above, is one of the earliest examples of
magnetic spiraling \cite{Meyer1956:JPR,Herpin1959:CR,Herpin1961:JPR} and may
provide better clues than FeAs as to how short-period spirals can form. The
spiral has an even shorter period than FeAs, the material is a metal, and,
unlike FeAs, Mn $d$-states are removed from the Fermi level, so the system may
be a better representation of a model with localized moments and an interaction
transferred \emph{via} itinerant electrons of a different nature. The magnetic
structure consists of ferromagnetic Mn planes (local moments are in-plane)
stacked along the crystallographic $c$ axis, with the in-plane magnetization
direction rotating from plane to plane. The rotation angle varies with
temperature, from 60$^{\circ }$ at $5\text{ K}$ to 40$^{\circ }$ at $250\text{
K}$ \cite{Nagata1999:JAC}. The N\'{e}el temperature is $T_{N}=363\text{ K}$
\cite{Meyer1956:JPR} and the material transitions from the spiral to a
ferrimagnetic fan-like structure at room temperature upon application of a $\sim
10\text{ kOe}$ magnetic field
\cite{Samata1998:JPCS,Meyer1956:JPR,Herpin1959:CR}, which gives rise to a giant
magnetoresistance effect \cite{Samata1998:JPCS}.

The crystal structure itself is also interesting. The Au atoms, which have 5
neighbors each, have much larger atomic radii when compared with Mn, and so they
form the framework that holds the structure together, with the Mn atoms fitting
into the center of cubic Au cages throughout the lattice. Such an arrangement
has implications for the electronic structure, which we will discuss later.

The common explanation for the MnAu$_{2}$ spiral is magnetic frustration, where
$\abs{J_{1}} < 4 \abs{J_{2}}$ \cite{Enz1961:JAP,Nagamiya1962:PTP}. In this
notation, $J_{1}$ is the exchange between nearest neighbor planes and $J_{2}$ is
the exchange between 2nd nearest neighbor planes. This interpretation was
supported by density functional (DFT) calculations \cite{Udvardi2006:PRB}, in
which the exchange constants were calculated using a relativistic extension
\cite{Udvardi2003:PRB} of the torque method \cite{Liechtenstein1987:JMMM} within
the screened Korringa-Kohn-Rostoker formalism \cite{Weinberger2000:CMS}, and it
was reported that $\abs{J_{1}} < 4 \abs{J_{2}}$ was satisfied. However, the
presence of highly itinerant carriers casts doubt on the idea of fully
describing the magnetism in this system using a $J_{1}-J_{2}$ Heisenberg model.
It is possible that the standard Heisenberg model can even fail to give a
qualitative description of the magnetic state, as in the case of the Fe-based
superconductors where such a description is dramatically inadequate, see
Refs.~\cite{Yaresko2009:PRB,Glasbrenner2014:PRB}. Furthermore, while
computational estimates for $J_{1,2}$ seem to satisfy the spiral criterion,
direct calculations (not performed in Ref.~\cite{Udvardi2003:PRB}) show that the
true DFT ground state is a uniform ferromagnet and not a spiral.

We are not aware of any other first-principles studies of the magnetic
interactions and the ground state of MnAu$_{2}$. Overall, first-principles
calculations of MnAu$_{2}$ have been sparse, aside from the above reference and
a pair of reports with calculations of the density of states
\cite{Hsu2006:JAC,Hsu2007:JAC}. It is worth revisiting this problem using
modern, full potential DFT calculations with noncollinear spin configurations
and extracting the exchange parameters from total energy calculations, rather
than by the torque perturbation theory with spherically-symmetrized potentials
\cite{Udvardi2006:PRB}. Given the good separation between the Mn and Au
electrons, one may hope to elucidate microscopic reasons for the spiral
ordering.

The paper is organized as follows. In Section \ref{sect-methods} we will detail
our computational methods for calculating the electronic structure and total
energy and extracting the exchange constants. We then follow in section
\ref{sect-discussion} with a report of our results and a subsequent discussion.
Our main result is that, contrary to Ref.~\cite{Udvardi2006:PRB}, a ``vanilla''
density functional theory does \emph{not} account for the spirals in MnAu$_{2}$.
However, upon accounting for on-site Coulomb correlations by applying an
LSDA$+U$ correction to the Mn $d$ orbital, we see that spiral solutions appear
for reasonable values of $U$ and agree with the helical angle from experiment.
This is an unexpected result, as the Hubbard $U$ enhances localization and
suppresses itinerant effects. Here, however, the correlated electrons forming
local moments are different from the itinerant electrons mediating the magnetic
interaction, and so a typical analysis using superexchange fails in such
materials. Instead, the magnetism needs to be reanalyzed in a way similar to
dilute magnetic semiconductors and Kondo lattices. We show, in particular, that
the main effect of the application of $U$ is to reduce hybridization between Mn
bands and Au electrons forming the Fermi surface. While the nearest neighbor
superexchange is suppressed as $1/U$, the transferred RKKY-type interaction goes
as $1/U^{2}$, introducing partial cancellation between the antiferromagnetic
superexchange and ferromagnetic transferred interaction between the neighboring
layer, which, in turn, enhances the $\abs{J_{2}/J_{1}}$ ratio. We describe our
conclusions in section \ref{sect-conclusions}.

\section{Computational methods}
\label{sect-methods}

We employed density functional theory (DFT) in three different implementations
to study spin spirals in MnAu$_{2}$. We used PAW potentials as implemented in
VASP \cite{vasp1,vasp2} and full potential linear augmented planewaves as
implemented in ELK \cite{elk} and WIEN2K \cite{wien2k}. The
Perdew-Burke-Ernzerhof generalized gradient approximation (GGA) \cite{pbe} was
used for the exchange-correlation functional in all three codes and the local
spin-density approximation (LSDA) \cite{lsda} was also used in ELK. Correlation
effects were considered in MnAu$_{2}$ using the DFT$+U$ method in the fully
localized limit \cite{Liechtenstein1995:PRB}, in which an empirical Hubbard $U$
is introduced on the $d$ orbitals of the Mn and/or Au atoms. Also, for a better
comparison with the atomic sphere approximation (ASA) calculations of
Ref.~\cite{Udvardi2006:PRB}, which were not a full potential treatment, we have
performed selected calculations in the ASA using a linear muffin-tin orbitals
(LMTO) code \cite{Andersen1975:PRB}.

The material MnAu$_{2}$ belongs to the space group $I4/mmm$ with Wyckoff
positions 2a for Mn and 4e for Au, which yields planes of Mn and Au atoms (2
layers of Au between each Mn layer). We set the lattice parameters to
$a=3.37013\text{ \AA }$ and $c=8.75894\text{ \AA }$ and the internal parameter
for Au to $z_{\text{Au}}=0.34$. The experimental ground state of MnAu$_{2}$ is a
spin spiral, where ferromagnetic Mn planes (local moments are oriented in-plane)
rotate about the $c-$axis with a noncollinear pitch vector close to the
incommensurate $\pi /2c$. To simulate the magnetic state, we consider spin
spirals in MnAu$_{2}$ with two different methods. The first is to construct
explicit spirals in supercells using noncollinear moments in the $xy$-plane
commensurate with $\mathbf{q}=\{0,0,\pi /2c\}$, which is done using the GGA
functional in both VASP and ELK. The second approach is to use a spin spiral
method to simulate spirals in a primitive cell with one Mn atom, which is done
using the LSDA functional in ELK alone.

We fit the energy calculations using the above methods to the
one-dimensional Hamiltonian, 
\begin{align}  
	\label{eq-hamiltonian} H &= \text{const.} + \sum_{i} J_{i} \cos \left( i \theta
	\right),
\end{align}
where the sum is taken over the layers of Mn atoms. Our primary interest is in
the ratio $\abs{J_{2}/J_{1}}$, so at a minimum we kept the first two terms in
the sum with constants $J_{1}$ and $J_{2}$, and then we varied the number of
layers in the sum to evaluate the robustness of the fit and the extracted
parameters.

\section{Discussion}
\label{sect-discussion}

\begin{figure}[ptb]
\begin{center}
\includegraphics[width=0.48\textwidth]{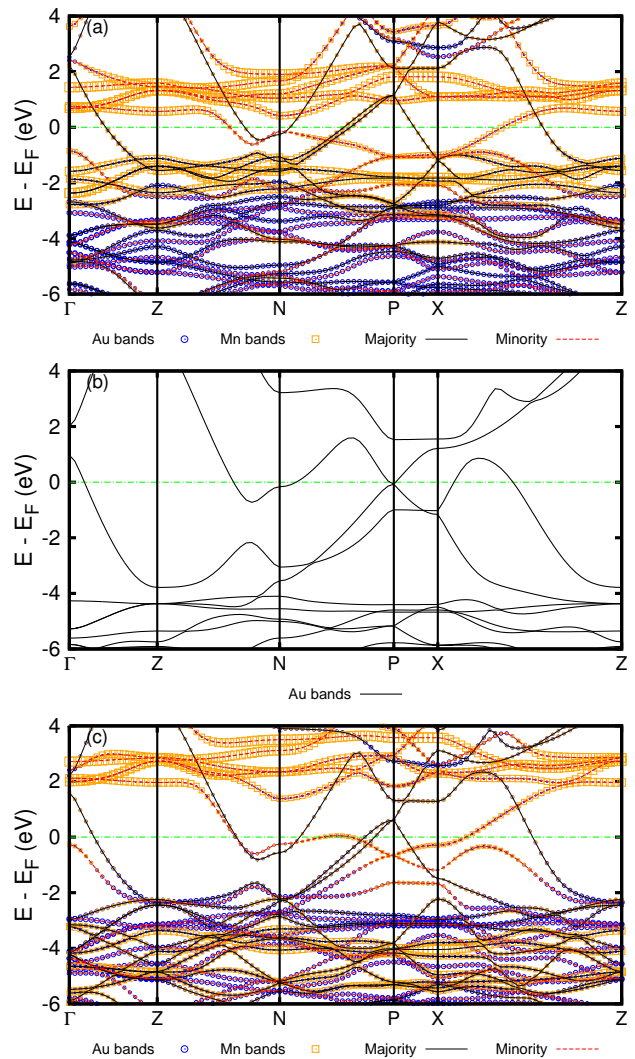}
\end{center}
\caption{(Color online) The spin- and species-resolved (see legend) band
structure of MnAu$_{2}$ calculated using the FLAPW code ELK. The plot point size
corresponds to the species weight. (a) The MnAu$_2$ band structure using the
LSDA functional. (b) Hypothetical Au-only band structure with Mn atoms removed
from the unit cell and Fermi level chosen to reflect Au$^{1-}$ charge transfer.
(c) The MnAu$_2$ band structure using the LSDA$+U$ functional with $U=4.7 \text{
eV}$.} \label{pic-bands}
\end{figure}

\begin{figure}[ptb]
\begin{center}
\includegraphics[width=0.48\textwidth]{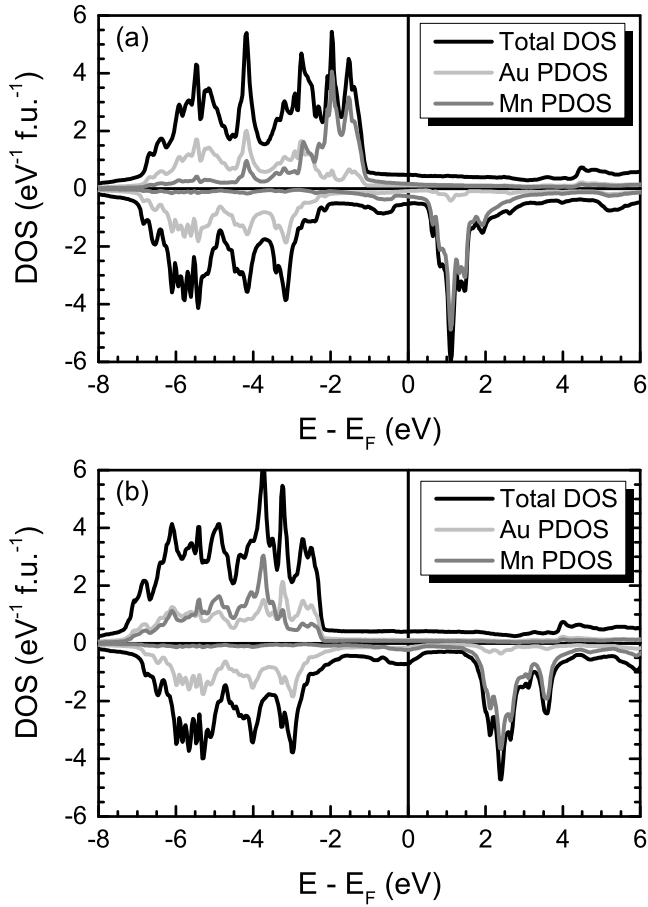}
\end{center}
\caption{The total and species-projected DOS of MnAu$_{2}$ calculated using the
FLAPW code ELK. (a) The DOS using the LSDA functional. (b) The DOS using the
LSDA$+U$ functional with $U=4.7 \text{ eV}$.} \label{pic-dos}
\end{figure}

We calculated the band structure and density of states (DOS) of ferromagnetic
MnAu$_{2}$ using the LSDA functional, shown in Figs.~\ref{pic-bands}(a) and
\ref{pic-dos}(a). The Mn $d$-bands are fully spin-split, corresponding to the
ionic configuration of Mn(d$^{5}$) and the formal Mn valency of $5+$. The total
moment of the system is 3.93 $\mu_{B}/\text{Mn}$, and the reduction of the
moment from the ideal 5 $\mu _{B}$ is due to the hybridization of Mn with Au.
The bands crossing the Fermi energy consist of both Mn and Au character that is
spin-dependent: the minority bands have $\sim 1.6$ times more Mn weight than the
majority bands, while the majority bands have $\sim 1.5$ times more Au weight
than the minority bands. We note that while the Au bands are polarized at the
Fermi energy, the net moment on the Au ions is zero.

The crystal structure of MnAu$_{2}$, as mentioned earlier, consists of Mn atoms
placed in cubic Au cages. The band structure and DOS suggest that the electronic
structure at the Fermi level is mostly determined by the Au atoms, therefore we
calculated the band structure of a hypothetical Au system where the Mn atoms
have been removed, shown in Fig.~\ref{pic-bands}(b). In the real system there is
a charge transfer of one electron per Au ion, so the Fermi level is chosen to
reflect an ionic charge of Au$^{1-}$. Comparing panels (a) and (b) shows the
remarkable similarity between the two band structures, indicating that much of
the electronic structure is due to the Au atoms only. The spin majority band
crossing the Fermi energy in panel (a) is the same Au band in panel (b), while
the spin minority band originates from the Mn atoms when they are placed into
the structure, shifting the Au bands upwards in energy.

\begin{figure}[ptb]
\begin{center}
\includegraphics[width=0.48\textwidth]{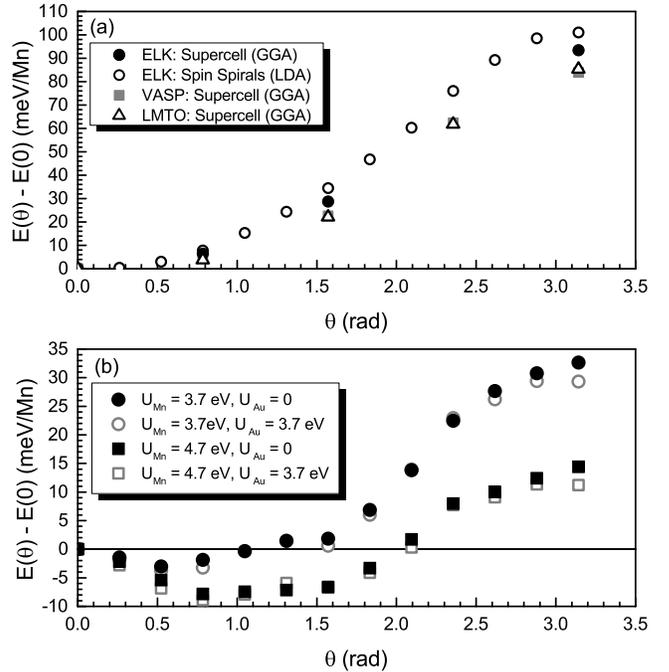}
\end{center}
\caption{The energy difference between a helical, in-plane spiral with angle
$\theta$ and a ferromagnetic configuration in MnAu$_{2}$. (a) The energy
dependence for LSDA and GGA functionals. The calculations are either spin spiral
calculations or explicit supercell calculations using VASP, ELK, or LMTO, see
the legend. (b) The energy dependence using LSDA$+U$, where the Hubbard $U$ is
applied to Au and/or Mn $d-$states. See the legend for the values of $U$.}
\label{pic-energycalcs}
\end{figure}

We next calculated the energy as a function of the spiral's helical angle
$\theta$ using the spin spiral method of ELK and explicit spirals in supercells
in both ELK and VASP. For the spin spiral method we used the LSDA functional
\cite{spinsprl} and for the supercell calculations we used the GGA functional.
We then computed the difference of $E(\theta)-E(0)$, comparing the energy of the
spiral state with the energy of the ferromagnetic configuration. The results are
summarized in Fig.~\ref{pic-energycalcs}(a). We consistently find in all cases
that there is a preference for the ferromagnetic ground state. The qualitative
trend across codes and functionals is the same. They differ with respect to the
energy of the antiferromagnetic configuration, with the LSDA spiral method
yielding the highest energy and the GGA VASP calculation yielding the lowest
energy. Overall, this shows that DFT does not support a stable spin spiral,
which is in disagreement with KKR calculations from Ref.~\cite{Udvardi2006:PRB}.

The origin of this disagreement cannot be ascribed to using different density
functionals, as we found the same result with LSDA and GGA, nor in the
approximations used to represent the crystal potential and electron (spin)
density. We have verified, using a LMTO method that employs the same spherical
approximation as the KKR method of Ref. \cite{Udvardi2006:PRB}, that the ground
state is still ferromagnetic, see Fig.~\ref{pic-energycalcs}(a). An important
difference may be that Ref.~\cite{Udvardi2006:PRB} uses the perturbative torque
method to calculate the exchange constants, while we performed total energy
calculations for explicit spin spiral configurations.

This results also hold when spin-orbit coupling (SOC) is turned on. The band
structure of MnAu$_{2}$ with and without SOC is practically identical even
though in principle Au is heavy enough to support non-negligible SOC effects.
Importantly, the moment on Au is zero and therefore relativistic magnetic
interactions of the DM type are excluded.

Although MnAu$_{2}$ is a good metal, the Mn $d-$states are quite localized and
are subject to local Hubbard correlations, not accounted for in straight DFT. It
is well known that other compounds with Mn$^{2+}$ require a Hubbard $U$ on the
order of $3-5 \text{ eV}$ to reproduce the correct positions of Mn bands. With
this in mind, we employed the LSDA$+U$ method in combination with the spin
spiral calculations in ELK to incorporate additional electronic correlations. We
applied the Hubbard $U$ to the Mn $3d$ orbitals, using two plausible values for
$U$, $U=3.7\text{ eV}$ and $U=4.7\text{ eV}$. The parameter $J$ was set to
$J=0.7\text{ eV}$. In addition, we also checked if the application of
$U=3.7\text{ eV}$ and $J=0.7\text{ eV}$ to the Au $5d$ orbitals affected the
results. The results of these calculations are in Fig.~\ref{pic-energycalcs}(b).
We found that spirals form when $U$ is applied to Mn states, while adding an
additional $U$ to Au had a negligible effect. For $U=3.7\text{ eV}$, a shallow
energy minimum of $\sim3\text{ mev}$ appears around $\theta=30^{\circ }$, and
for a $U=4.7\text{ eV}$ a deeper well of $\sim8\text{ meV/Mn}$ appears around
$\theta=45^{\circ}$. Finally, we note that including $U$ compresses the overall
energy scale compared with the results of Fig.~\ref{pic-energycalcs}(a).

This evolution can be understood if we picture the relevant magnetic interaction
as a combination of the nearest-neighbor-plane antiferromagnetic superexchange,
proportional to $t_{\perp }^{2}/\Delta _{\uparrow \downarrow}$, and the
transferred magnetic interaction mediated by the Au electrons. Note that
$t_{\perp}$ is the effective interplane hopping and $\Delta_{\uparrow \downarrow
}$ is the energy cost of transferring a Mn $d$ electron to a neighboring site
and flipping its spin. The transfer interaction can be visualized as the spin
susceptibility of the Au subsystem multiplied by the square of the Mn-Au
interaction vertex. In our case this vertex is, to a first approximation, the
Schrieffer-Wolff interaction \cite{Cibert:2008,Khomskii:2010}, which is
proportional to $\tau ^{2}/\Delta _{sd}$, where $\tau$ is the Mn-Au hopping
amplitude and $\Delta_{sd}$ measures how removed the occupied Mn $d$-states are
from the Fermi level. For a Fermi energy that falls roughly in the middle of the
lower and upper Mn Hubbard bands, $\Delta _{sd} \approx \Delta _{\uparrow
\downarrow }/2$. Furthermore, in the LSDA$+U$ calculations $\Delta _{\uparrow
\downarrow } \approx \Delta_{ex}+U,$ where $\Delta _{ex}$ is the LSDA (Stoner)
exchange splitting.

The transferred interaction is distance dependent and is allowed to change sign,
in contrast to the superexchange interaction. In particular, our calculations
are consistent with a ferromagnetic nearest-neighbor interaction and an
antiferromagnetic next-nearest neighbor interaction. The net nearest-neighbor
exchange parameter is $J_{1} \approx J_{\text{SE}} + J_{\text{Mn-Au}}^{(1)}$,
where $J_{\text{Mn-Au}}^{(1)} < 0$ and $\abs{J_{\text{SE}}} <
\abs{J_{\text{Mn-Au}}^{(2)}}$, while $J_{2} \approx J_{\text{Mn-Au}}^{(2)}$,
where $J_{\text{Mn-Au}}^{(2)} > 0$. Since $J_{\text{SE}} \propto
1/\Delta_{\uparrow \downarrow}$ and $J_{\text{Mn-Au}} \propto 1/\Delta_{\uparrow
\downarrow}^{2}$, as $U$ increases both $J_{1}$ and $J_{2}$ shall decrease, but
$J_{1}$ will decrease more rapidly as $J_{\text{SE}}$ starts to play a more
dominant role.

This is precisely what we observe in the calculations. Fitting to
Eq.~\ref{eq-hamiltonian} we were able to reliably extract the constants $J_{1}$
and $J_{2}$, which remained robust regardless of the number of additional
neighbors we included in the fit. The extracted constants are summarized in
Table \ref{table-jconstants} along with the constants from
Ref.~\cite{Udvardi2006:PRB}. In all cases the $J_{1}$ parameter is ferromagnetic
and $J_{2}$ is antiferromagnetic. Consistent with the argument above, the
nearest-neighbor exchange $J_{1}$ is very sensitive to correlations, decreasing
by a factor of $\sim 3$ for $U=3.7\text{ eV}$ and $\sim 6$ for $U=4.7\text{
eV}$. In contrast, the decrease for $J_{2}$ is moderate for $U=3.7\text{ eV}$
and very small when $U$ is increased further to $U=4.7\text{eV}$. Our LSDA value
of $J_{2}$ is in good agreement with Ref.~\cite{Udvardi2006:PRB}, while we find
$J_{1}$ to be larger by a factor of 2.

\begin{table}[ptb]
\caption{The extracted exchange constants obtained by fitting the results of
Fig.~\ref{pic-energycalcs} to Eq.~\ref{eq-hamiltonian}. For comparison, the
calculated constants from Ref.~\cite{Udvardi2006:PRB} are also included. All
constants are reported in units of meV.} \label{table-jconstants}
\begin{tabular}{|c|c|c|c|c|}
\hline
Const. & LSDA & \multicolumn{2}{|c|}{LSDA$+U$} & LSDA \\ \hline
&  & $U = 3.7 \text{ eV}$ & $U = 4.7 \text{ eV}$ & Udvardi \emph{et.~al.} \\ 
\hline
$J_{1}$ & -49.37 & -16.50 & -8.51 & -25.16 \\ \hline
$J_{2}$ & 8.17 & 6.43 & 6.30 & 8.31 \\ \hline
\end{tabular}%
\end{table}

The band structure for MnAu$_{2}$ with $U=4.7\text{ eV}$ is shown in
Fig.~\ref{pic-bands}(c) and the DOS is in Fig.~\ref{pic-dos}(b). As described,
the primary effect of introducing $U$ is to move the Mn $d$ bands away from the
Fermi level, which decreases the hybridization of Au with Mn, localizing Mn and
decreasing the energy difference between FM and layered AFM configurations. This
in turns lowers the polarization of the Au bands, which mediate the transferred
interaction, which then weakens the ferromagnetic exchange of $J_{1}$. We
conclude that the spiral instability in MnAu$_{2}$ is driven by correlation
effects which enhance the importance of the superexchange interaction relative
to the transferred interaction induced by Au polarization.

Finally, we tried to identify the microscopic reason for the transferred
interaction to change sign between $c/2$ and $c$ along the $c-$axis. A natural
explanation would be in terms of a Fermi surface nesting at an appropriate
$q_{z}$. We note that $q_{z}$ does not have to coincide exactly with the spiral
wave vector; adding superexchange will shift it towards shorter wavelengths.
Unfortunately, one cannot use the nonmagnetic MnAu$_{2}$ for investigating
nesting effects, as the Mn moments are very large and a linear response
treatment is not appropriate. The hypothetical (Au$^{1-}$)$_{2}$ system is more
appropriate due to the similarity of the band structures in
Figs.~\ref{pic-bands}(a) and \ref{pic-bands}(b), which become increasingly more
alike with increasing $U$. To check for nesting, we calculated the
noninteracting one-electron susceptibility for (Au$^{1-}$)$_{2}$, defined as
\begin{align*}
	\chi_{0}(\mathbf{q}) &=
	\sum_{\alpha,\beta,\mathbf{k}} \frac{f(\epsilon_{\alpha,\mathbf{k}}) -
	f(\epsilon_{\beta,\mathbf{k}+\mathbf{q}})}{\epsilon_{\beta,\mathbf{k} +
	\mathbf{q}} - \epsilon_{\alpha,\mathbf{k}} + i \gamma}.
\end{align*}
Note that this expression neglects any matrix elements arising from the fact
that Au wave functions deviate from plane waves. We found a weak maximum at
$\mathbf{q}=\{0,0,2\pi /c\}$, which would support antiferromagnetically aligned
layers and would induce $J_{1} > 0$ and $J_{2} < 0$, opposite to what is needed
for the experiment and what has been derived in Table \ref{table-jconstants}
from the calculations. As it stands, further progress in understanding the
Au-mediated transferred interaction requires calculations of the full spin
susceptibility, which includes matrix elements and Stoner renormalization. At
the present moment we do not have the capability to perform these calculations.

\section{Conclusion}
\label{sect-conclusions}

We have revisited the origin of spirals in MnAu$_{2}$ using accurate, full
potential, noncollinear DFT calculations, and found that contrary to previous
findings, DFT alone is not sufficient to sustain a helical spiral state. We find
that the spirals in MnAu$_{2}$ are supported by Hubbard correlations, which
localize the Mn $d$ electrons and strongly reduce the ferromagnetic coupling
between neighboring Mn layers. This mechanism is in contrast to the common
origin of spirals, which are typically due to relativistic effects such as the
DM interaction or one-electron effects such as Fermi surface nesting. This
uncommon physical origin may be present in other materials where traditional
explanations of spirals fail, such as FeAs.

\begin{acknowledgments}
We are very grateful to Kay Dewhurst for his invaluable help and advice in
setting up and performing spiral calculations in ELK. I.I.M. acknowledges
Funding from the Office of Naval Research (ONR) through the Naval Research
Laboratory's Basic Research Program. J.K.G. acknowledges the support of the NRC
program at NRL.
\end{acknowledgments}

\bibliography{mnau2}

\begin{thebibliography}{40}
\bibitem [{spi()}]{spinsprl}%
  \BibitemOpen
  \href@noop {} {}\bibinfo {note} {Our use of the LSDA functional over the GGA
  functional is a practical one. The spin spiral algorithm in ELK works well
  with the LSDA functional, but there are numerical issues when the GGA
  functional is used. For example, with GGA $\mathbf{q} = \{0,0,0\}$ and
  $\mathbf{q}=\{0,0,1\}$ yield different energies, even though they should be
  equivalent.}\BibitemShut {Stop}%
\end{thebibliography}

\end{document}